\documentclass[proof]{WileyASNA-v1}

\articletype{Article Type}%

\received{26 August 2024}

\begin{document}

\title{Probing the propeller regime with wide neutron star binaries}

\author[1]{M.D. Afonina }

\author[2]{S.B. Popov*}

\authormark{AFONINA AND POPOV}

\address[1]{\orgdiv{Sternberg Astronomical Institute}, \orgname{Lomonosov Moscow State University}, \orgaddress{\state{Moscow}, \country{Russia}}}

\address[2]{ \orgname{International Center for Theoretical Physics}, \orgaddress{\state{Trieste}, \country{Italy}}}

\corres{*Sergei B. Popov, Strada Costiera 11, 34151, Trieste, Italy. \email{sergepolar@gmail.com}}


\abstract{We model the duration of the propeller stage in wide binary systems with neutron stars and calculate the time of accretion onset for various propeller models. We apply our modeling to the symbiotic X-ray binary SWIFT J0850.8-4219. Unless a propeller with a very slow spin-down is operating, 
it is very improbable to find a system similar to SWIFT J0850.8-4219 at the propeller stage. 
 Then we model the evolution of a neutron star in a binary with a solar-like companion. 
We calculate for which orbital separations and magnetic fields a neutron star can start to accrete while the companion is still on the Main sequence. 
We demonstrate that for the magnetic field $B\lesssim10^{12}$~G neutron stars at the orbital separation $a\gtrsim 1$~AU do not reach the propeller stage. In the case of a slow propeller spin-down, neutron stars never start to accrete. For the more rapid propeller spin-down, 
a neutron star can start to accrete or spend a long time at the propeller stage depending on the parameters. 
 
}




\maketitle


\section{Introduction}\label{intro}

 Early studies of X-ray binaries were focused on systems with small orbital separations as they demonstrate large luminosity due to high accretion rates. 
A large amount of gravitationally captured matter (when it is not super-Eddington) allows for some simplifications in treating the evolution of such systems. 
 In wide binaries, accretion proceeds due to the stellar wind capture. Thus, the luminosity is usually relatively low even for massive donors, well below the Eddington limit. For very wide systems or/and for low-mass Main sequence (MS) donors the accretion is potentially possible but the corresponding luminosity can be below the present-day limit of detectability. 

 Wide binary systems with low accretion rates or with dormant compact objects open new interesting possibilities to study the accretion and evolution of neutron stars (NSs). 
 A low supply of external material results in a slower evolution of an NS. Thus, it takes longer for the compact object to reach the stage of accretion. In particular, the propeller stage can be sufficiently long. This makes wide binaries an interesting site to probe the properties of this elusive stage. 

 Recently, mainly due to the data collected by the {\it Gaia} mission, several binary systems with yet directly undetected compact objects have been discovered, see e.g., \cite{2024OJAp....7E..58E, 2024A&A...686A.299S}. 
 Apparently, black holes and NSs in these systems are non-accreting as no X-ray emission is detected. Still, the situation is unclear as the upper limits do not contradict accretion at low rates. It is important to understand the present-day state and evolution of NSs in such systems. 

 In this paper, we present modeling of the evolution of the symbiotic X-ray binary (SyXB) SWIFT J0850.8-4219. For this system, it was suggested by \cite{2024MNRAS.528L..38D} 
that it could be at the propeller stage. We analyze the probability of this scenario for several models of the propeller spin-down. Basic assumptions about the magneto-rotational evolution are given in Sec.~2. Our initial results on this source were presented by \cite{2024Univ...10..205A}. Here in Sec.~3, we show the calculations made with slightly different assumptions. 
Then, in Sec.~4, we use the same set of propeller models to model the evolution of a wide NS binary with a solar-like donor. 
In Sec.~5 we briefly discuss our results and present our conclusions.

\section{Magneto-rotational evolution of neutron stars}\label{evol}

In this section, we introduce some standard characteristic quantities used to describe the magneto-rotational evolution of NSs. Details can be found, e.g. in \cite{1992ans..book.....L, 2024Galax..12....7A}. 

At first, we remind the reader of several characteristic radii. 
\begin{itemize}
    \item
The gravitational capture radius (aka Bondi radius):
    $R_{\text{G}}={2 G M}/{v^2}.$
Here $G$ is the Newton constant, $M=1.4\,M_\odot$ is the NS mass, and $v$ is the NS velocity relative to the surrounding medium.
 \item
The light cylinder radius:
     $R_\text{l}=c/\omega.$
 Here $c$ is the speed of light and $\omega=2\pi /P$ is the spin frequency. 
 \item
The Alfven radius (which can be different from the magnetospheric radius $R_\mathrm{m}$): $R_A = [ { \mu^2}/({2 \Dot{M} \sqrt{2GM}})  ]^{2/7}.$
Here $\mu=B R_\mathrm{NS}^3$ is the magnetic moment, $B$ is the surface magnetic field, and $R_\mathrm{NS}=10^6$~cm is the NS radius.
$\dot M$ is the accretion rate. 
 \item
 Corotation radius: $R_{\text{co}} = ( {GM}/{\omega^2})^{1/3}.$
Note, that usually this quantity is used as the radius of the centrifugal barrier. 
However, here following \cite{2023MNRAS.520.4315L} we approximate this limit as
 $R_\text{cb} = 0.87R_\text{co}$.
 \item 
Finally, we define the radius where the external pressure equalizes the pressure of the relativistic particle wind from the pulsar. This is so-called Shvartsman radius:
 $R_{\text{Sh}}=[  {(\xi \mu^2 \omega^4)}/{(4 \pi  c^4 \rho v^2)} ]^{1/2}$, where $\xi=2$. 
This equation is valid for the case when $R_\mathrm{Sh}>\mathrm{max}(R_\mathrm{G}, R_\mathrm{l})$. Here $\rho$ is the density of the surrounding medium and $\xi$ is a numerical factor. 
\end{itemize}

During its evolution, an NS can pass the following stages: ejector (E), propeller (P), accretor (A), 
 see Table~\ref{fig1}. In this paper, we ignore the georotator stage. 
We also add an intermediate stage of a transient ejector (TE). In the case $R_\mathrm{G}>R_\mathrm{l}$ the classical ejector stage is expected to end when $R_\mathrm{Sh}=R_\mathrm{G}$.
However, sometimes the usual (supersonic) propeller stage cannot be established immediately. An intermediate stage might appear.  Due to intensive energy release at the magnetospheric boundary, a rarefied envelope with a shallow pressure profile forms around the NS. This opens the possibility for an intermediate transient ejector regime. Energy and angular momentum losses at this stage are uncertain. However, due to a relatively low density at the magnetospheric boundary, we do not expect significant spin-down due to the transient propeller mechanism. We assume that the average spin-down rate at this stage is approximately equal to the ejector spin-down rate. This stage lasts till $R^\mathrm{env}_\mathrm{Sh}>R_\mathrm{l}$ in the rarified envelope with pressure $\sim r^{-3/2}$. Here the modified Shvartsman radius is calculated as $ R^\mathrm{env}_\mathrm{Sh}$ (Table~\ref{tab2}).
 Conditions for a transition from stage to stage are given in Table~\ref{tab1}.

At the propeller stage, the magnetosphere radius $R_\text{m}$ (Table~\ref{tab2}) is calculated as follows: first, we assume that $R_\text{m} < R_\text{G}$. If the result exceeds $R_\text{G}$, then we recalculate it given $R_\text{m} > R_\text{G}$.

Spin-down is calculated as:
    $I ({d\omega}/{dt}) = -K.$
Here $I=10^{45}$~g~cm$^2$ is the moment of inertia and $K$ is the braking torque different at various stages.  For the ejector and transient ejector we use $K_\mathrm{E}=\xi \mu^2/R_\mathrm{l}^3$. 
Braking torques for different propeller models are given in Table~\ref{tab2}. At the accretor stage, either $K_\text{A} =k_\mathrm{t}\mu^2/R_\mathrm{cb}^3 - \dot{M} \eta \Omega R_\text{G}^2$, where $k_\mathrm{t}\approx0.4$, $\eta=1/4$, $\Omega$ is the orbital frequency, or $K_\text{A} = k_\mathrm{t}\mu^2/R_\mathrm{cb}^3 - \dot{M} \sqrt{GMR_\text{A}}$ if the disc is formed (i.e. $\sqrt{GMR_\text{A}} \leq \eta \Omega R_\text{G}^2$).
Other details of our model can be found in \citep{2024Univ...10..205A}.

\begin{center}
\begin{table}[t]%
\centering
\caption{Conditions for the transition between evolutionary stages. For the transition to occur, a radius of interaction between the outer matter and the pulsar wind ($R_\text{Sh}$ or $R_\text{Sh}^\text{env}$) or the NS magnetosphere ($R_\text{m}$ or $R_\text{A}$) must be equal to another critical radius, depending on the stage. \label{tab1}}%
\tabcolsep=0pt%
\begin{tabular*}{20pc}{@{\extracolsep\fill}ll@{\extracolsep\fill}}
\toprule
\textbf{Stages} & \textbf{The condition of transition}  \\
\midrule
Ejector-TE & $R_{\text{Sh}} \le \text{max}(R_\text{G},~R_\text{l})$ \\ 
TE-Propeller & $R_\text{Sh}^\text{env} \le \text{min}(R_\text{G},~R_\text{l})$ \\ 
Propeller-Accretor & $R_\text{m} \le R_\text{cb}$  \\
\bottomrule
\end{tabular*}
\begin{tablenotes}
\end{tablenotes}
\end{table}
\end{center}


\begin{center}
\begin{table*}[t]
\caption{The propeller models, corresponding spin-down torques $K$, magnetospheric radii $R_\text{m}$ and Shvartsman radii in the envelope $R_\text{Sh}^\text{env}$, which are the characteristic radii of interaction between gravitationally captured material and the relativistic particle wind of the NS.
\label{tab2}}
\centering
\renewcommand{\arraystretch}{1.4}
\begin{tabular*}{500pt}{@{\extracolsep\fill}lcccc@{\extracolsep\fill}}
\toprule
\textbf{Model} & \textbf{The braking torque $K$} & \textbf{$R_\text{m}$, if $R_\text{m} < R_\text{G}$} & \textbf{$R_\text{m}$, if $R_\text{m} > R_\text{G}$} & \textbf{$R_\text{Sh}^\text{env}$} \\ \midrule
A1 & $\Dot{M} \omega R_{\text{m}}^2$ & $[{\mu^2 v R_\text{G}^{1/2}}/{(2 \dot{M} \omega^2)}]^{2/13}$ & $[{\mu^2 v R_\text{G}^{2}}/{(2 \dot{M} \omega^2)}]^{1/8}$ & $R_\text{G}[c^4\dot{M}v/(\xi\mu^2\omega^4)]^2$ \\ \midrule
A & $\Dot{M} \omega R_{\text{m}}^2$  & \multirow{4}{*}{$R_\text{A}^{7/9}R_\text{G}^{2/9}$} & \multirow{4}{*}{$[{\mu^2 R_\text{G}^{2}}/{(2 \dot{M} v)}]^{1/6}$} & \multirow{4}{*}{$R_\text{G}[\xi\mu^2\omega^4/(c^4\dot{M}v)]^2$} \\ \cline{1-2}
B & $\dot{M} \sqrt{2GMR_{\text{m}}}$ &  & \\ \cline{1-2}
C  & $\dot{M} \text{max}(v^2,~v^2_{\text{ff}}(R_\text{m})) / (2\omega)$ &  &  \\ \cline{1-2}
D  & $\dot{M} v^2 / (2\omega)$ &  & \\ \bottomrule
\end{tabular*}
\begin{tablenotes}
\item The authors of the models: A1, A~--- \cite{1975SvAL....1..223S}, B~--- \cite{1973ApJ...179..585D}, C~--- \cite{1975AA....39..185I}, D~--- \cite{1981MNRAS.196..209D}.  
\end{tablenotes}
\end{table*}
\end{center}

\section{Symbiotic X-ray binary SWIFT J0850.8-4219}\label{swift}

In this section, we calculate the spin evolution of the NS in a wide binary with a massive companion that does not fill its Roche lobe. Following the suggestion by \cite{2024MNRAS.528L..38D}, we discuss the conditions necessary for an NS in a SyXB to be observed at the propeller stage.

\subsection{Observational data}\label{data}
SyXBs form a small group of X-ray binaries consisting of an accreting NS and a late-type giant. \cite{2019MNRAS.485..851Y} suggest that the number of these systems in the Galaxy is only 40-50. In the majority of SyXBs, the donor star is a low-mass giant. However, there are also a few systems that contain a red supergiant, for instance, 4U 1954+31 \citep{2020ApJ...904..143H} and SWIFT J0850.8-4219 \citep{2024MNRAS.528L..38D}.

In SWIFT J0850.8-4219 the donor star is a red supergiant with a mass $\sim10\div20 \, M_\odot$  and an effective temperature $T_\text{eff}=3820\pm 100$~K. The orbital period of the system and NS spin period are not known. 
\citeauthor{2024MNRAS.528L..38D} exclude that the primary component can be a black hole, as the X-ray spectrum is hard: $N(E) \propto E^{-\Gamma}$, where the photon index is $\Gamma<1$. The X-ray luminosity of the system is $L=(4\pm1) \times 10^{35}$~erg~s$^{-1}$. This is too high for accretion onto a white dwarf, but too low for a typical accreting NS. The authors consider a wind accretion without the  Roche lobe overflow and propose that the system contains an NS at the propeller stage. Thus, only a small part of material accretes onto the NS surface, which is the reason for the X-ray luminosity reduction.

\subsection{Modeling and results}\label{results}

As in our previous study \citep{2024Univ...10..205A}, we consider a model that represents a SyXB similar to SWIFT J0850.8-4219. For the evolution of the donor, we utilize the PARSEC track \citep{2015MNRAS.452.1068C} for a $M_*=14\,M_\odot$ star to obtain its radius $R_*(t)$, mass $M_*(t)$, effective temperature $T_*(t)$, and mass loss rate $\dot{M}_\text{w}(t)$. We assume that the time needed for the primary component to become an NS is at least $\sim7$~Myr and use only the part of the track that starts at $t=7$~Myr. We consider a circular orbit with a semi-major axis $a=1280\,R_\odot$, so there is no Roche lobe overflow. 


\cite{2000A&A...362..295V} show that the wind velocity profile of the O and B stars follows a beta-type law with $\beta = 1$. We use the same law with $\beta=2$ for the red supergiant phase \citep{2010ASPC..425..181B}:
    $v_\text{w}(r) = v_\infty\left(1-{R_*}/{r}\right)^\beta$.
Here $v_\infty$ is the terminal velocity, which is proportional to the escape velocity $v_\text{esc}=\sqrt{GM_*/R_*}$. The wind of the secondary component undergoes two bi-stability jumps, which are the consequence of changes in $T_*$. This results in sharp jumps of $\dot{M}_\text{w}$ 
and influences the terminal velocity: $v_\infty/v_\text{esc}=2.6,\,1.3,\,0.7$ before, between and after the two bi-stability jumps \citep{1999A&A...350..181V}. The second jump is associated with the onset of the red supergiant phase.

The density of the wind is determined from the continuity equation
    $\rho (r) = \dot{M}_\text{w} / (4\pi r^2 v_\text{w})$.
The relative velocity $v$ includes the velocity of the compact object in the frame of reference of the donor $v = \sqrt{v_\text{w}^2+G(M+M_*)/r}$.
Finally, with known $v$ and $\rho$ on the NS orbit $r=a$ we can determine an accretion rate 
    $\dot{M} \approx 2.5\pi(GM)^2 v^{-3} \rho$.
 The coefficient $2.5$ is in agreement with the numerical solutions for moderate Mach numbers \citep{1997A&A...320..342F, 2024ApJ...966..103P}.

\begin{figure}[t]
\centerline{\includegraphics[width=78mm]{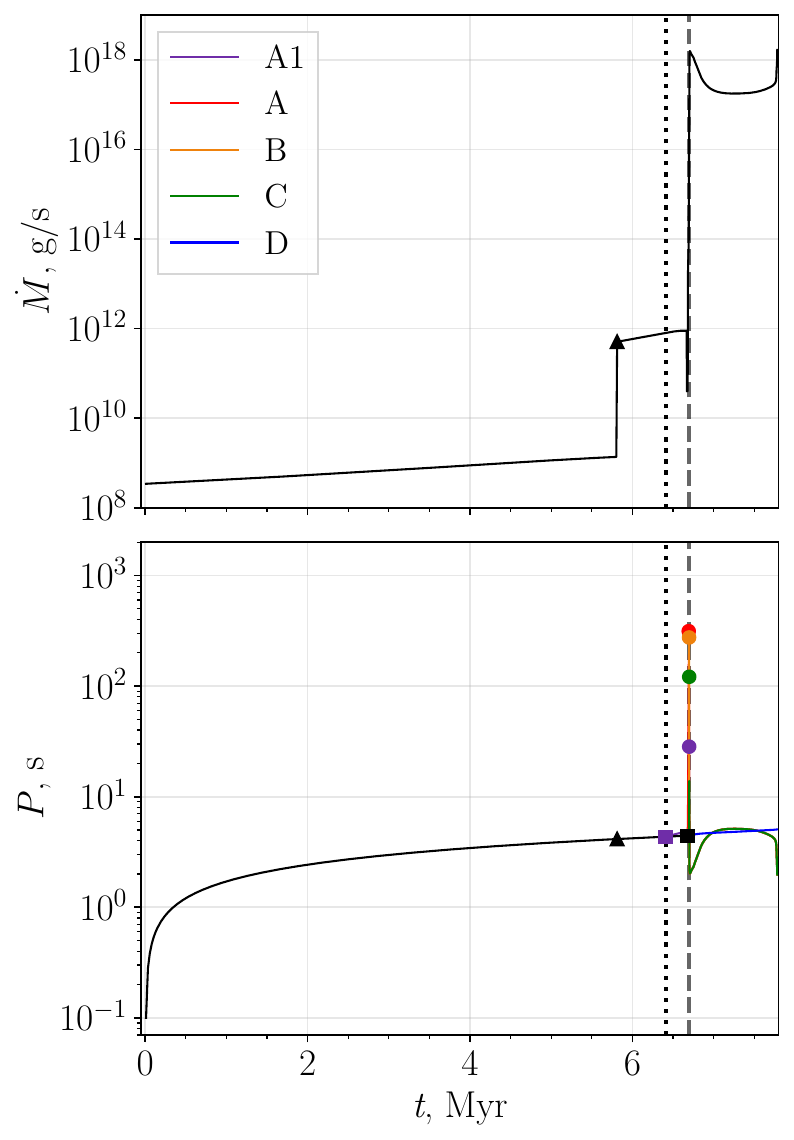}}
\caption{The rate of matter capture $\dot{M}$ (top panel) and the spin evolution of the NS with the constant magnetic field $B=4\times10^{12}$~G and the circular orbit $a=1280\,R_\odot$ around a star $14\,M_\odot$ (bottom panel). Different colors correspond to different propeller models. All NSs are born as ejectors. The spin period $P$ at the ejector stage and the plot of $\dot{M}$ are colored black as they are identical to all propeller stages. The black-filled triangle marks the transition to the TE stage. The purple-filled square and the black dashed vertical line indicate the transition to the propeller stage in model A1. The black-filled square and the grey dashed line show the transition to the propeller stage for the rest of the models. The transition to the accretor stage occurs almost simultaneously for all propeller models and coincides with the grey dashed line. The colored dots indicate the propeller-accretor transition spin periods. All NSs, except for the propeller model D, reach the accretor stage.
\label{fig1}}
\end{figure}

The accretion rate $\dot{M}$ and the spin period of an NS with $B=4\times10^{12}$~G in the SyXB are presented in Fig.~\ref{fig1}. The main driver of the spin evolution of the NS is the evolution of the donor star. Thus, most transitions between NS stages occur due to the sharp increase in the accretion rate.

The evolution of NSs in models A, B, and C proceeds as follows. First, the NS is born as an ejector. After the first bi-stability jump at $5.8$~Myr the TE stage starts, as the external pressure increases and the material becomes gravitationally captured by the NS. Then the NS remains to spin down mainly due to the pulsar losses, which is independent of $\dot{M}$. The propeller stage starts at the second increase in $\dot{M}$ at $6.7$~Myr. With the accretion rate $10^{18}$~g~s$^{-1}$ the propeller spin-down is so efficient that the NS reaches the propeller-accretor transition period in less than 20 thousand years. The matter falling onto the NS has a large angular momentum, so a disk around the NS is formed. After the onset of accretion, the spin-up torque of the disk balances the spin down at the accretor stage, so, the spin period of the NS is equal to the equilibrium spin period.

In model A1, the NS evolves similarly with only one difference~--- due to a distinct envelope structure the TE stage ends earlier. 
In model D, the NS does not reach the accretor stage. Therefore, among the propeller models considered, only the model with the most inefficient spin-down mechanism, model D, provides a significant probability of observing a system like SWIFT J0850.8-4219 with an NS at the propeller stage.

These new results does not change the main conclusions of our previous work \citep{2024Univ...10..205A} where we did not apply the TE stage and assumed $R_\text{cb}=R_\text{co}$. 

\section{Wide non-interacting binaries with neutron stars discovered by Gaia}\label{gaia}

 In this section, we present our preliminary calculations of an NS evolution in a binary system with a solar-like companion. 
 Such systems are now discovered by {\it Gaia}, see a review in \cite{2024NewAR..9801694E}. 

\subsection{Observational data}\label{gaia_data}

 Astrometric data obtained by {\it Gaia} provide the possibility to identify binary systems with invisible compact companions. In particular, with NSs. 
 Up to now, several tens of candidates have been identified, e.g. \cite{2022arXiv220700680A, 2024OJAp....7E..58E}. In some cases, mass determinations do not exclude the white dwarf nature of invisible objects. Still, in most cases, compact companions might be NSs. 
 It is expected that when the fourth data release ({\it Gaia} DR4) appears in 2026, many tens of new systems of this kind will be discovered. Thus, it is timely to discuss the evolution of NSs in such binaries in detail as future observations might allow probing properties of propeller and accretor stages for low accretion rates. 

 {\it Gaia} is mainly sensitive to systems with orbital periods $\sim 1\div 3$~yrs. Indeed, most of the systems under discussion have periods in this range \citep{2024A&A...686A.299S, 2024OJAp....7E..58E}. 
In addition, some short-period binaries, $P_\mathrm{orb}\lesssim 1^\mathrm{d}$, are found \citep{2022ApJ...940..165Y, 2023ApJ...944L...4L} 
 due to spectral observations with {\it LAMOST}. For one of them -- 1527+3536, -- a weak X-ray counterpart is known \citep{2024A&A...686A.299S}. However, in this paper, we do not discuss such binaries as their evolution might include at least one episode of intense interaction (maybe, with a common envelope formation). Thus, the model we use here is not applicable to them.

 Masses of optical companions of the systems reported by \cite{2024A&A...686A.299S, 2024OJAp....7E..58E} are in the range $\sim 0.7\div 1.3 \, M_\odot$. Thus, the stars are solar-like. For our illustrations, we consider a system with one solar mass companion. This allows us to apply detailed calculations of the wind evolution performed for our Sun. 

 As we a dealing with wide binaries with a light normal companion and an NS, we expect orbits to be eccentric due to the kick and mass ejection during a supernova explosion. Indeed, most of the systems presented by \cite{2024A&A...686A.299S, 2024OJAp....7E..58E} have eccentricities $e\sim 0.3-0.7$. However, here we use a circular orbit as it helps to illustrate the properties of the evolution more clearly. 
 We plan to analyze eccentric binaries in more detail in our future studies.





\subsection{Modeling and results}\label{gaia_results}

We consider a long-term evolution of an NS in a system with a solar-like secondary component, while the donor remains an MS star, i.e., the first $\sim10$~Gyr.




For the solar wind evolution we adopt one of the models considered by \cite{2015A&A...577A..28J}. According to them, for $t>300$~Myr, $\dot{M}_\text{w} = \dot{M}_{\text{w}0} (4.6\text{~Gyr}/t)^{0.75}$ where $\dot{M}_{\text{w}0}=2\times10^{-14}\,M_\odot/$yr. For the stars younger than $300$~Myr, $\dot{M}_\text{w}$ remains constant. In this model, $v_\infty$ does not change over time. We are interested in wide orbits where $v_\text{w}\approx v_\infty$, i.e. $a\gtrsim0.2$~AU. Therefore, for our calculations, we use a single value of the wind velocity, $v_\text{w} = 400$~km~s$^{-1}$ (the slow component of the wind, see \citealt{2005ApJ...623..511A}). Then, the relative velocity is $v = \sqrt{v_\text{w}^2+G(M+M_\odot)/r}$. 
We determine the density on the NS orbit as $\rho = \dot{M}_\text{w}/(4\pi r^2 v_\text{w})$ and finally calculate an accretion rate $\dot{M} = 4\pi(GM)^2 v_\text{w}^{-3} \rho$, assuming that $v$ is considerably higher than the sonic speed.

\begin{figure}[t]
\centerline{\includegraphics[width=78mm]{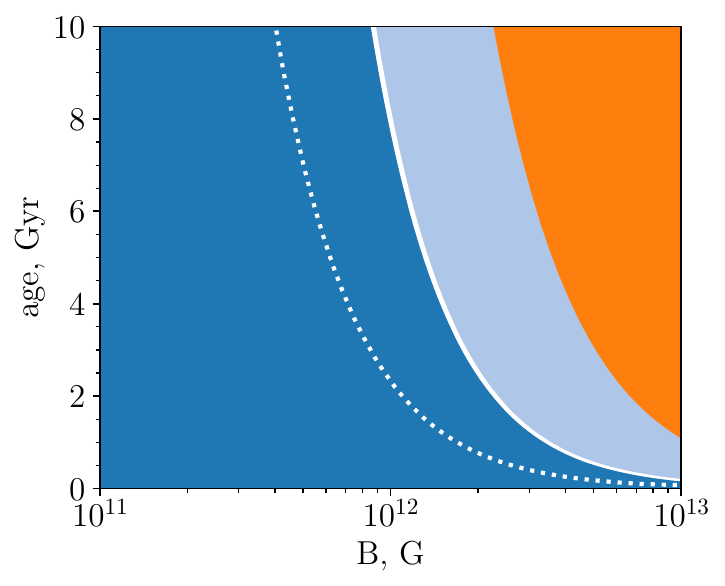}}
\caption{The evolutionary stages of NS with magnetic field $B$ in a circular orbit with semi-major axis $a=1$~AU at a given age. The ejector stage is dark blue. The dotted line shows the transition to the TE stage. The short propeller stage in model A is shown in white. For model B, the propeller stage is both the white and light blue regions. The rest of the plot is the accretor stage (the light blue and orange areas in model A and orange in model B).
\label{figB}}
\end{figure}

\begin{figure}[t]
\centerline{\includegraphics[width=78mm]{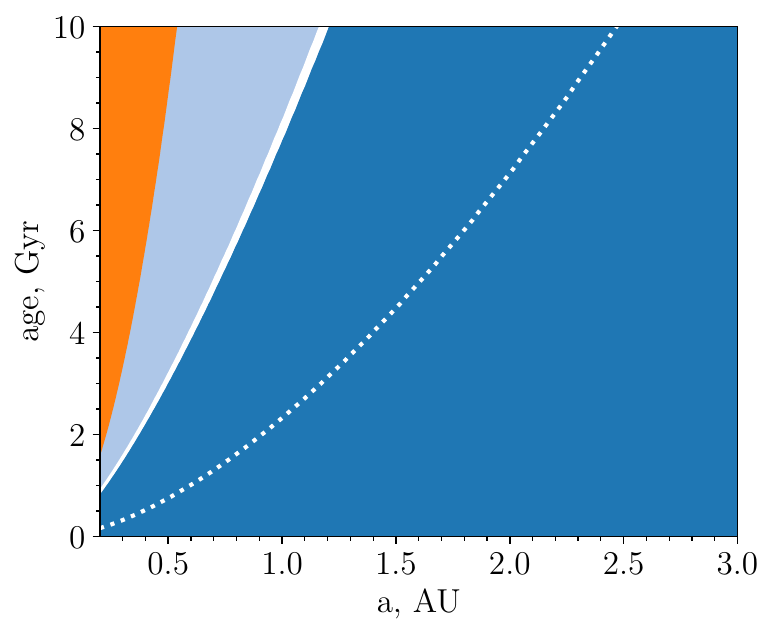}}
\caption{The evolution of an NS with the constant magnetic field $B=10^{12}$~G on a circular orbit with the semi-major axis $a$. The colors and the white dotted line represent the evolutionary stages for the propeller models A and B as in Fig.~\ref{figB}. 
\label{figa}}
\end{figure}

The evolutionary stages of NSs with different constant magnetic fields in the case of $a=1$~AU are illustrated in Figure~\ref{figB}. Figure~\ref{figa} shows the NS with the standard magnetic field value $B=10^{12}$~G in various circular orbits. The range of $a$ is in accordance with the data by \cite{2024OJAp....7E..58E}. In order to demonstrate the onset of accretion, we have selected only propeller models A and B. Model A1 would result in a plot that is very similar to model A. The propeller spin-down mechanism in models C and D is too ineffective to bring the NS to the accretor stage. 

We set the initial spin period as $P_0=100$~ms. Thus, all considered NSs are born at the ejector stage. 
The value of $\dot{M}$ does not influence the ejector spin-down. However, a lower external pressure requires the NS to achieve a longer spin period for the transition. So, NSs farther from the secondary component spend more time as ejectors. 
The ejector spin-down rate strongly depends on $B$. So, the NS with the higher magnetic field evolves faster. Additionally, the decreasing dependence $\dot{M}_\text{w}(t)$ makes the ejector stage even longer for the low-field NS. The same arguments apply to the duration of the subsequent TE stage. Thus, NSs with the magnetic fields $\lesssim 10^{12}$~G on orbits $a\gtrsim 1$~AU can not reach the accretor stage regardless of the propeller model we apply.

The propeller stage in model A has a high spin-down rate. This makes the propeller stage to be $\lesssim500$~Myr for all considered parameters. Therefore, the onset of accretion in NSs with $B\gtrsim10^{12}$~G and $a\lesssim1$~AU is possible. 
In model B, the duration of the propeller stage is comparable to the ejector stage duration. The propeller spin-down visibly depends on $\dot{M}$ and $B$ and has a higher rate with higher values of these parameters. In model B, in order to reach the accretor stage, the NS should either be closer than $0.5$~AU to the secondary component or keep a strong magnetic field $\gtrsim 3\times10^{12}$~G on a large timescale.

As a result, an NS with a magnetic field $\gtrsim10^{12}$~G located sufficiently close to the second star ($a\lesssim1$~AU) may have a significant probability of being observed as an accretor if the spin-down mechanism at the propeller stage is efficient enough.

\section{Discussion and conclusion}\label{disc}

In this note, we presented a modification of the model of the evolution of a symbiotic X-ray binary SWIFT J0850.8-4219 and preliminary results of the evolution of an NS in a wide binary system with a solar-like companion. For various assumptions about the propeller stage, we calculated its duration and the time of accretion onset.

 The duration of the propeller stage can be very different in different scenarios. Only in the case of a very slow propeller spin-down, 
like in the model by \cite{1981MNRAS.196..209D}, 
there is a significant probability of finding a system similar to SWIFT J0850.8-4219 at the propeller stage. Otherwise, accretion starts quickly as soon as the mass loss from the companion increases. 
Thus, symbiotic X-ray binaries with well-determined parameters can be a good probe for the propeller stage evolution.

 Several tens of candidates for wide NS binaries with solar-like companions are now detected by {\it Gaia}. 
We calculated for which orbital separations and magnetic fields an NS can start to accrete while the companion remains on the Main sequence. 
We demonstrated that for the magnetic field $B\lesssim10^{12}$~G NSs at the orbital separation $a\gtrsim 1$~AU do not reach the propeller stage. In the case of a slow propeller spin-down, NSs never start to accrete. For the more rapid propeller spin-down, like those proposed by \cite{1975SvAL....1..223S} 
and \cite{1973ApJ...179..585D}, 
an NS can start to accrete or spend a long time at the propeller stage depending on the parameters. 

The calculations of the NS evolution in a wide binary system presented above are done here for the case of circular orbits and constant magnetic fields. Our modeling for decaying fields and eccentric binaries will be presented elsewhere.

In many respects, NS properties in wide binaries found in the {\it Gaia} data might be similar to long-sought isolated accreting NSs, e.g. \cite{2015MNRAS.447.2817P} and references therein. 
As isolated accretors remain undiscovered, studies of NSs in wide binaries now provide the best opportunity to understand the properties of the propeller stage at low accretion rates.



\section*{Acknowledgments}

S.B.P. thanks the Organizers and participants of the  `XMM-Newton Workshop 2024'. 
 M.D.A. was supported by the Basis Foundation.



\subsection*{Conflict of interest}

The authors declare no potential conflict of interest.


\nocite{*}
\bibliography{propeller_XMM_popov}%



\end{document}